\documentclass[twocolumn,aps,prd,superscriptaddress,preprintnumbers,nofootinbib,10pt]{revtex4-2}
\usepackage{multirow}
\usepackage{amsmath}
\usepackage{amssymb}
\usepackage[dvipdf,dvips]{graphicx}
\usepackage{color}
\usepackage{hyperref}
\usepackage{url}
\usepackage{slashed}
\usepackage{subfigure}
\usepackage[usenames,dvipsnames]{xcolor}
\usepackage{amsmath}
\usepackage{amsfonts}
\usepackage{float} 
\usepackage{amssymb}
\usepackage{epsfig}
\usepackage{graphics}
\usepackage{euscript}
\usepackage{slashed}
\usepackage{epstopdf}
\usepackage[utf8]{inputenc}
\allowdisplaybreaks
\usepackage[normalem]{ulem}
\usepackage{pifont}
\usepackage{dsfont}
\usepackage{MnSymbol}
\usepackage{graphicx}
\usepackage{latexsym}
\usepackage{tikz-feynman}
\usepackage{easyReview}

\begin{document}

	\title{Using Weyl operators to study  Mermin's  inequalities in Quantum Field Theory}

	\author{P. De Fabritiis} \email{pdf321@cbpf.br} \affiliation{CBPF $-$ Centro Brasileiro de Pesquisas Físicas, Rua Dr. Xavier Sigaud 150, 22290-180, Rio de Janeiro, Brazil}
	
	\author{F. M. Guedes} \email{fmqguedes@gmail.com} \affiliation{UERJ $–$ Universidade do Estado do Rio de Janeiro,	Instituto de Física $–$ Departamento de Física Teórica $–$ Rua São Francisco Xavier 524, 20550-013, Maracanã, Rio de Janeiro, Brazil}
	
	\author{M. S.  Guimaraes}\email{msguimaraes@uerj.br} \affiliation{UERJ $–$ Universidade do Estado do Rio de Janeiro,	Instituto de Física $–$ Departamento de Física Teórica $–$ Rua São Francisco Xavier 524, 20550-013, Maracanã, Rio de Janeiro, Brazil}
	
	\author{I. Roditi} \email{roditi@cbpf.br} \affiliation{CBPF $-$ Centro Brasileiro de Pesquisas Físicas, Rua Dr. Xavier Sigaud 150, 22290-180, Rio de Janeiro, Brazil}
	
	\author{S. P. Sorella} \email{silvio.sorella@gmail.com} \affiliation{UERJ $–$ Universidade do Estado do Rio de Janeiro,	Instituto de Física $–$ Departamento de Física Teórica $–$ Rua São Francisco Xavier 524, 20550-013, Maracanã, Rio de Janeiro, Brazil}

	\begin{abstract}

Mermin's inequalities are investigated in a Quantum Field Theory framework by using von Neumann algebras built with Weyl operators. We devise a general construction based on the Tomita-Takesaki modular theory and use it to compute the vacuum expectation value of the Mermin operator,  analyzing the parameter space and explicitly exhibiting a violation of Mermin's inequalities. Therefore, relying on the power of modular operators, we are able to demonstrate that Mermin's inequalities are violated when examined within the vacuum state of a scalar field theory.

	\end{abstract}

	\maketitle

\section{Introduction}

The study of entanglement within the realm of Quantum Field Theory poses a formidable challenge~\cite{Witten18,Hollands18}. The combination of Quantum Mechanics and Special Relativity leads to a very rich and sophisticated framework, where questions such as locality and causality come up as a building block~\cite{Weinberg05,Duncan12}. A Quantum Field Theory naturally presents quantum fields as operator-valued distributions and a vacuum state that unveils a profoundly intricate structure, enlightened by the Reeh-Schlieder theorem~\cite{Haag92}. Therefore, dealing with entanglement in Quantum Field Theory demands sophisticated tools such as Algebraic Quantum Field Theory and the Tomita-Takesaki modular theory~\cite{Summers87a, Summers87b}.

Although the idea of entanglement~\cite{Horodecki09} was introduced in the 1930s~\cite{EPR35}, it was only with the groundbreaking work of Bell, Clauser, Horne, Shimony, and Holt (Bell-CHSH)~\cite{Bell64,CHSH69, Clauser74,Clauser78} that this counterintuitive feature of nature could be experimentally confirmed~\cite{Freedman72,Aspect82,Zeilinger98}. Entanglement can be considered the deepest departure from classical physics contained in Quantum Mechanics~\cite{Werner01}, and its existence is now beyond any doubt, with the violation of Bell-CHSH inequalities already confirmed in many contexts through highly sophisticated experiments in different physical systems~\cite{Giustina13,Giustina15,Hensen15,Zhong19,Storz23}. 

The Bell-CHSH inequalities, although mainly investigated in the context of Quantum Mechanics, can also be  investigated within the realm of Quantum Field Theory~\cite{Summers87a,Summers87b} and are currently receiving much attention~\cite{Haar23,BRST23,Weyl23,Sorella23,Sorella22}. Many experimental tests in the high-energy context were proposed recently~\cite{Afik21,Fabbrichesi21,Severi22,Afik22,Barr22,Afik23,Ashby23}, which will allow us to investigate entanglement in a regime never explored before. Remarkably, the ATLAS collaboration announced very recently the first observation of quantum entanglement between a pair of quarks and the highest-energy measurement of entanglement to date~\cite{ATLAS}.

Quantum entanglement can be generalized for systems with more than two subsystems~\cite{Greenberger07,Dur00,Briegel01}. Multipartite entanglement was experimentally observed for the first time in~\cite{Bouwmeester99} and can be considered a relevant resource for quantum information theory~\cite{McCutcheon16}. Mermin proposed a generalization of Bell inequalities for multipartite systems~\cite{Mermin90}, which received further developments in the sequence~\cite{Ardehali92,Roy91,Belinskii93,Gisin98}. It is well known that GHZ-type states~\cite{GHZ90} maximally violate Mermin inequalities, which have also been investigated for $W$-type states~\cite{Swain18} (see also~\cite{Mermin2}). Violations of Mermin inequalities were reported in many works, as one can see, for instance, in Refs.~\cite{Pan00,Erwen14,Zhao03,DiCarlo10,Neeley10}. Moreover, Mermin inequalities violations for superconducting qubits using a quantum computer were reported in~\cite{Alsina16a}.

This work aims to provide a framework for investigating the violation of Mermin inequalities~\cite{Mermin90} within the vacuum state of a real scalar field in the $(3+1)$-dimensional Minkowski spacetime, thereby extending our previous findings related to the Bell-CHSH inequality~\cite{Weyl23}. To that end, we shall rely on unitary Weyl operators which, due to their algebraic properties, prove to be very helpful for the study of this class of inequalities within the domain of Quantum Field Theory. We remark that Mermin inequalities have already
been investigated in the Quantum Field Theory framework for Dirac spinor fields~\cite{Mermin23}. 

The paper is organized as follows. Section~\ref{TF} is devoted to the presentation of a few basic concepts concerning real scalar fields, Weyl operators, von Neumann algebras and the Tomita-Takesaki modular theory. For the benefit of the reader, in Section~\ref{BTT} we briefly review how the aforementioned setup applies to the case of Bell-CHSH inequalities. Section~\ref{M} delves into the generalization of the previous construction to the case of Mermin inequalities. We will show that, when probed in the vacuum state of a relativistic Quantum Field Theory, these inequalities violate the bound expected from the local realism. The use of  Weyl operators will enable us to make a bridge with the recent discussion on the use of normal operators in Quantum Theory, as outlined in Section~\ref{N}. Finally, we state our conclusions in Section~\ref{C}.

\section{Theoretical framework}\label{TF}

Let us consider a free massive scalar field in a $(3+1)$-dimensional Minkowski space, with action given by
\begin{align} 
S = \int \!\! d^4x \left[\frac{1}{2} \left(\partial_\mu \phi\right)^2  - \frac{m^2}{2} \phi^2\right].  
\end{align}
The scalar field can be expanded in terms of creation and annihilation operators as 
\begin{equation} \label{qf}
\phi(t,{\vec x}) = \int \!\! \frac{d^3 {\vec k}}{(2 \pi)^3} \frac{1}{2 \omega_k} \left( a_k e^{-ikx}  + a^{\dagger}_k e^{ikx}  \right), 
\end{equation} 
with $\omega_k=k^0=\sqrt{{\vec{k}}^2 + m^2}$. The canonical commutation relations for these creation and annihilation operators read
\begin{align}\label{ccr}
[a_k, a^{\dagger}_q] &= (2\pi)^3 2\omega_k \delta^3({\vec{k} - \vec{q}}), \\ \nonumber 
[a_k, a_q] &= [a^{\dagger}_k, a^{\dagger}_q] = 0, 
\end{align}
Using the above definitions, one  obtains the commutator between the scalar fields for arbitrary  spacetime points:
\begin{align} 
\left[ \phi(x) , \phi(y) \right] = i \Delta_{PJ} (x-y), \label{caus} 
\end{align}
where the  Pauli-Jordan distribution is defined by:
\begin{align}\label{PJ}
	 i \Delta_{PJ}(x-y) \!=\!\! \int \!\! \frac{d^4k}{(2\pi)^3} \varepsilon(k^0) \delta(k^2-m^2) e^{-ik(x-y)},
\end{align}
with $\varepsilon(x) \equiv \theta(x) - \theta(-x)$. From the above expression, one can see that the Pauli-Jordan distribution is Lorentz invariant and odd under the exchange $(x-y) \rightarrow (y-x)$. Moreover,  it vanishes outside of the light cone, ensuring that measurements at space like separated points do not interfere. As such,  the Pauli-Jordan distribution encodes the information about  locality and relativistic causality.

The quantum fields are operator-valued distributions and must be smeared to give well-defined operators acting on the Hilbert space~\cite{Haag92}. That is, we can use a real smooth test function with compact support $h(x) \in \mathcal {C}_{0}^{\infty}(\mathbb{R}^4)$ to define the smeared quantum field
\begin{align} \label{smqf}
\phi(h) = \int \! d^4x \; \phi(x) h(x).
\end{align}
Plugging Eq.~\eqref{qf} into Eq.~\eqref{smqf}, we can rewrite the smeared quantum field  as $\phi(h) = a_h + a^{\dagger}_h$, defining the smeared version of the creation and annihilation operators by
\begin{align} 
a_h &= \int \frac{d^3 {\vec k}}{(2 \pi)^3} \frac{1}{2 \omega_k}  {\hat h}^{*}(\omega_k,{\vec k}) a_k, \nonumber \\
a^{\dagger}_h &= \int \frac{d^3 {\vec k}}{(2 \pi)^3} \frac{1}{2 \omega_k} {\hat h}(\omega_k,{\vec k}) a^{\dagger}_k, 
\end{align} 
with ${\hat h}(p) = \int \! d^4x \; h(x)  e^{ipx}$ for the Fourier transform of $h(x)$.
Using the smeared fields, one introduces  the  Lorentz-invariant inner product in the space of the test functions with compact support, given by the smeared version of the Wightman two-point function, namely,
\begin{align} \label{IP}
    \langle f \vert g \rangle &= \langle 0 \vert \phi(f) \phi(g) \vert 0 \rangle \nonumber \\
    &= \frac{i}{2} \Delta_{PJ}(f,g) + H(f,g) \nonumber \\
    &= \int \!\! \frac{d^3p}{(2\pi)^3} \frac{1}{2 \omega_p} f^*(p) g(p)  
\end{align}
where $ \Delta_{PJ}(f,g)$ is the smeared version of Eq.~\eqref{PJ} and $H(f,g)$ is the symmetric combination of the smeared field product. Thus, we can rewrite the commutator~\eqref{caus} in its smeared version as $\left[\phi(f), \phi(g)\right] = i \Delta_{PJ}(f,g)$. We remark that, upon using the canonical commutation relations and the above-defined inner product~\eqref{IP}, one  finds that
\begin{align}
    \left[a(f), a^\dagger(g)\right] = \langle f \vert g \rangle.
\end{align}

Let $O$ be an open set in  Minkowski spacetime, and let $M(O)$ be the space of test functions belonging to  $\mathcal{C}_{0}^{\infty}(\mathbb{R}^4)$ with support contained in $O$, that is,
\begin{align} 
	M(O) = \{ f \, \vert supp(f) \subset O \}. \label{MO}
\end{align}
We proceed by defining  the causal complement $O'$ of the spacetime region $O$ as well as the symplectic complement $M'(O)$ of the set $M(O)$ as 
\begin{align} 
O' &= \{ y \, \vert (y-x)^2 < 0, \; \forall x \in O \}, \nonumber \\	
 M'(O) &= \{ g \, \vert  \Delta_{PJ}(g,f) =0, \; \forall f \in M(O) \}. \label{MpO}
\end{align}
With the above definitions, causality can be rephrased by stating that $\left[ \phi(f), \phi(g) \right] = 0$ whenever    $f \in M(O)$ and $g \in M'(O)$. Furthermore, we can also rephrase locality using the expression $M(O') \subset M'(O)$ (see Refs.~\cite{Summers87a,Summers87b}).

Now, let us introduce the Weyl operators~\cite{Weyl23}, a well-known class of unitary operators given by
\begin{align}
    W_f = e^{i \phi(f)}.
\end{align}
These operators give rise to the so-called Weyl algebra
\begin{align}
    W_f W_g = e^{-\frac{i}{2} \Delta_{PJ}(f,g)} \,W_{f+g}.
\end{align}
One can  see that when the supports of $f$ and $g$ are spacelike separated, the Pauli-Jordan distribution vanishes, and thus the Weyl operators behave as $W_f W_g = W_{f+g} = W_g W_f$. It should also be stressed that, unlike the quantum field $\phi(f)$,  the Weyl operators  are bounded. Upon using $\phi(f) = a_f + a^\dagger_f$, one can compute the vacuum expectation value of the Weyl operator, finding that
\begin{align}
    \langle 0 \vert W_f \vert 0 \rangle = e^{-\frac{1}{2} \vert\vert f \vert\vert^2},
\end{align}
where $\vert\vert f \vert\vert^2 = \langle f \vert f \rangle$ and $\vert 0 \rangle$ is the  Fock vacuum.

One can define the observable algebra $\mathcal{A}(O)$ associated with the spacetime region $O$ as the von Neumann algebra obtained by taking products and linear combinations of the Weyl operators defined on $M(O)$~\footnote{For the definitions adopted here, see the appendix of Ref.~\cite{Weyl23}.}. It is known that, by the Reeh-Schlieder theorem, the vacuum state $\vert 0 \rangle$ is both cyclic and separating for the von Neumann algebra $\mathcal{A}(O)$~\cite{Witten18, Haag92}. We can therefore make use of the powerful Tomita-Takesaki modular theory \cite{Bratteli97} and introduce the  antilinear unbounded operator $S$ acting on the von Neumann algebra $\mathcal{A}(O)$ as
\begin{align} 
	S \; a \vert 0 \rangle = a^{\dagger} \vert 0 \rangle, \qquad \forall a \in \mathcal{A}(O),  \label{TT1}
\end{align}  
from which it follows that $S^2 = 1$ and $S \vert 0 \rangle = \vert 0 \rangle$. Performing the polar decomposition of the operator $S$ \cite{Bratteli97}, one gets
\begin{align}
S = J  \Delta^{1/2}, \label{PD}    
\end{align} 
where $J$ is antiunitary  and $\Delta$ is positive and self-adjoint. These modular operators satisfy the following properties~\cite{Bratteli97}: 
\begin{align} 
	J \Delta^{1/2} J &= \Delta^{-1/2}, \quad \,\,	\Delta^\dagger = \Delta, \nonumber \\
	S^{\dagger} &= J \Delta^{-1/2},  \,\,\,\,\, J^{\dagger} = J, \nonumber \\
	\Delta &= S^{\dagger} S,  \quad \,\,\,\,\,\, J^2 = 1. \label{TTP}
\end{align}
According to the Tomita-Takesaki theorem~\cite{Bratteli97} it turns out that $J \mathcal{A}(M) J = \mathcal{A}'(M)$, that is, upon conjugation by the operator $J$, the algebra $\mathcal{A}(M)$ is mapped onto its commutant $\mathcal{A'}(M)$ 
\begin{equation} 
\mathcal{A'}(M) = \{ \; a' \, \vert \; [a,a']=0, \forall a \in \mathcal{A}(M) \;\}. \label{commA}
\end{equation} 
The Tomita-Takesaki construction has far-reaching consequences when it is applied to Quantum Field Theory. As far as the Bell inequalities are concerned, it gives a way of constructing Bob's  operators from Alice's by making use  of the modular conjugation $J$. That is, given Alice's operator $A_f$, one can assign the operator $B_f = J A_f J$ to Bob, with the guarantee that they commute with each other since by the Tomita-Takesaki theorem the operator  $B_f = J A_f J$ belongs to the commutant $\mathcal{A'}(M)$~\cite{Weyl23}.

When equipped with the Lorentz-invariant inner product $\langle f \vert g\rangle$ defined in~Eq.\eqref{IP}, the set of test functions gives rise to a complex Hilbert space $\mathcal{F}$ which enjoys many interesting properties, as outlined in \cite{Rieffel77}. It turns out that the subspaces $M$ and $iM$ are standard subspaces for $\mathcal{F}$, meaning that i) $M \cap i M = \{ 0 \}$ and ii) $M + i M$ is dense in $\mathcal{F}$. Moreover, as proven in~\cite{Rieffel77}, for standard subspaces it is possible to set a modular theory analogous to that of the Tomita-Takesaki theorem. One introduces an operator $s$ acting on $M + iM$ as
\begin{align}
    s (f+ih) = f-ih \;, \label{saction}
\end{align}
for $f,h \in M$. Notice that with this definition, it follows  that $s^2 = 1$. Using the  polar decomposition, one has:  
\begin{align}
    s = j \delta^{1/2},
\end{align}
where $j$ is an antiunitary operator and $\delta$ is  positive and self-adjoint.  Similarly to the operators $(J,\, \Delta)$, the  operators $(j,\, \delta)$ fulfill  the following properties:
\begin{align}
    j \delta^{1/2} j &= \delta^{-1/2}, \,\,\,\,\,\,  \delta^\dagger = \delta\nonumber \\
    s^\dagger &= j \delta^{-1/2}, \,\,\, j^\dagger = j \nonumber \\
    \delta &= s^\dagger s, \,\,\,\,\,\,\,\,\,\,\, j^2=1.
\end{align}
An important result \cite{Rieffel77} concerning the operator $s$ is the following: a test function $f$ belongs to $M$ if and only if 
\begin{equation} 
s f = f \;. \label{sff}
\end{equation}
In fact, to illustrate the reasoning behind this assertion, suppose that $f \in M$. On general grounds, owing to Eq.~\eqref{saction}, one writes 
\begin{equation}
sf = h_1 + i h_2 \;, \label{pv1}
\end{equation}
for some $(h_1,h_2)$. Since $s^2=1$ it follows that 
\begin{equation} 
f = s(h_1 + i h_2) = h_1 -i h_2  \;, \label{pv2}
\end{equation} 
so that $h_1=f$ and $h_2=0$. In much the same way, one has that $f' \in M'$ if and only if $s^{\dagger} f'= f'$. 

As shown in~\cite{Eckmann73},  the action of the operators $(j,\delta)$ on the von Neumann algebra $\mathcal{A}(M)$ is defined through 
\begin{align} 
 J e^{i {\phi}(f) } J  = e^{-i {\phi}(jf) }, \quad \Delta e^{i {\phi}(f) } \Delta^{-1} = e^{i {\phi}(\delta f) }. \label{jop}
\end{align} 
Also, it is worth noting that $f \in M \implies jf \in M'$. This property follows from 
\begin{equation} 
s^{\dagger} (jf) = j \delta^{-1/2} jf = \delta f = j (j\delta f) = j (sf) = jf \;. \label{jjf} 
\end{equation} 
We end this section by mentioning that, in the case of wedge regions in Minkowski spacetime, the spectrum of $\delta$ coincides with the positive real line, {\it i.e.}, $\log(\delta) = \mathbb{R}$ \cite{Bisognano75}, being an unbounded operator with a continuous spectrum.

\section{Bell-CHSH inequalities and the Tomita-Takesaki construction}\label{BTT}

Before addressing the issue of the Mermin inequalities, for the benefit of the reader, we briefly remind how the use of Weyl operators and Tomita-Takesaki modular theory leads to a rather simple setup for studying the Bell-CHSH inequality within the Quantum Field Theory framework. For more details, see Ref.~\cite{Weyl23}. 

To construct the Bell-CHSH inequality, we introduce Alice's operators as $(W_f, W_{f'})$. Moreover, using the modular conjugation $j$,  Bob's operators are given by $(W_{jf}, W_{jf'})$. From the discussion of the previous section, the operators $(W_{jf}, W_{jf'})$ turn out to commute with $(W_f, W_{f'} )$, as required by the Bell-CHSH inequality. Therefore, we can write the Bell-CHSH correlator~as
\begin{align}
\langle 0 \vert C \vert0 \rangle  =  \langle 0 \vert ( W_{f} +  W_{f'}  )  W_{jf} +  
 ( W_{f} -  W_{f'}  )  W_{jf'}\vert 0 \rangle.
\end{align}
Using the properties of Weyl operators outlined in the previous section, one obtains  
\begin{align}
\langle 0 \vert {C} \vert0 \rangle &=  e^{-\frac{1}{2} \vert\vert f+jf \vert\vert^2} +  e^{-\frac{1}{2} \vert\vert f'+jf \vert\vert^2} \nonumber \\
&+  e^{-\frac{1}{2} \vert\vert f+jf' \vert\vert^2} - e^{-\frac{1}{2} \vert\vert f'+jf' \vert\vert^2} \;. \label{chatres}
\end{align}
In order to evaluate the norms present in the above expression, we follow the procedure outlined in~\cite{Summers87a,Summers87b}. Using the fact that the operator $\delta$ has a continuous spectrum coinciding with the positive real line, we pick up the spectral subspace specified by $[\lambda^2-\varepsilon, \lambda^2+\varepsilon ] \subset (0,1)$. Let $\phi$ be a normalized vector belonging to this subspace. One notices that $j\phi$ is orthogonal to $\phi$, {\it i.e.}, $\langle \phi |  j\phi \rangle = 0$. In fact, from 
\begin{align} 
\delta^{-1} (j \phi) =  j (j \delta^{-1} j) \phi = j (\delta \phi), \label{orth}
\end{align}
it follows that the modular conjugation $j$ exchanges the spectral subspace $[\lambda^2-\varepsilon, \lambda^2+\varepsilon ]$ for $[1/\lambda^2-\varepsilon,1/ \lambda^2+\varepsilon ]$. Therefore, proceeding as in \cite{Weyl23}, we set 
\begin{align}
f  &= a (1+s) \phi \\
f'  &= a' (1+s) i\phi \;, \label{nmf}
\end{align}
where $(a,a')$ are arbitrary real constants. Since $s^2=1$, it turns out that 
\begin{equation} 
s f = f \;, \qquad s f' = f'  \;, \label{sff}
\end{equation}
so that  both $f$ and $f'$ belong to $M$. Recalling that $\phi$ belongs to the spectral subspace $[\lambda^2-\varepsilon, \lambda^2+\varepsilon ] $, it follows that \cite{Weyl23}, 
\begin{align}
\vert\vert f \vert\vert^2  &= \vert\vert jf \vert\vert^2 = a^2 (1+\lambda^2) \nonumber \\
\vert\vert f' \vert\vert^2  &= \vert\vert jf' \vert\vert^2 = a'^{2} (1+\lambda^2) \;. \label{sfl}
\end{align}
One can also find the nonvanishing inner products
\begin{align} 
\langle f \vert jf \rangle &= 2 a^2 \lambda, \nonumber \\
\langle f' \vert jf' \rangle &= 2 a'^2 \lambda \;. \label{scalp} 
\end{align}
Therefore, using the above expressions, we can compute all the norms appearing in Eq.~\eqref{chatres}, finally obtaining for the Bell-CHSH inequality
\begin{align}
\langle C \rangle =  \left[ e^{-a^2 (1+\lambda)^2 } -  e^{-a'^2 (1+\lambda)^2 } +  2e^{-\frac{1}{2} (a^2 + a'^2) (1+\lambda^2)}  \right]. \label{fbchsh}
\end{align}
As shown in \cite{Weyl23}, the above simple expression is already able to capture the violation of the Bell-CHSH inequality. An interesting feature of Eq.~\eqref{fbchsh} is that it has been obtained by making direct use of the unitary Weyl operators only, {\it i.e.}, $W_f =   e^{i \phi(f)}$, without the need of introducing any Hermiticity procedure. This property will give us the opportunity to make a bridge with the recent ongoing discussion on the role of the normal operators, as advocated for in  \cite{Hu_2017,Erhard_2020}. 	

\section{Generalizing the results to the Mermin inequalities}\label{M}

We are now ready to generalize the previous setup to the case of Mermin's inequalities. Mermin's polynomials can be defined in a recursive manner according to the following rule~\cite{Alsina16a,Alsina16b}:
	\begin{align}\label{key}
		M_n = \frac{1}{2} M_{n-1} \left(A_n + A'_n\right) + \frac{1}{2} M'_{n-1} \left(A_n - A'_n\right).
	\end{align} 
where $A_n$ and $A'_n$ are dichotomic quantities that can take values $\pm 1$. In the above expression, the $M_n'$ operators can be obtained from $M_n$ by performing the changes $A_n \rightarrow A_n'$ and $A_n' \rightarrow A_n$. Here we will adopt $M_1 = 2 A_1$ as the first term in the recursive procedure. Following this definition, the maximum value that can be obtained in Quantum Mechanics is given by the following upper bound
\begin{align}\label{key}
		\vert \langle M_n \rangle \vert \leq 2^{\left(\frac{n+1}{2}\right)}.
\end{align}
For example, the third order Mermin's polynomial is given by
\begin{equation}
		M_3 = A' B C + A B' C + A B C' - A' B' C'. \label{third}
\end{equation}
Considering the absolute value, local realistic theories  obey the bound $\vert \langle M_3 \rangle_{\text{Cl}} \vert \leq 2 $. For  Quantum Mechanics, on the other hand, the upper bound is given by $\vert \langle M_3 \rangle_{\text{QM}}\vert \leq 4$. In the following we shall consider as an explicit example the case of the polynomial $M_3$, with the generalization to $M_n$ being straightforward. 

From the Quantum Field Theory point of view, $(A,A')$, $(B,B')$, and $(C,C')$ are pairs of bounded localized field operators with norm $\le 1$. The different pairs are meant to be spacelike separated. As done in the Bell-CHSH inequality case, these operators are built employing the unitary Weyl operators already introduced. One speaks of a Mermin inequality violation in the vacuum state whenever we have
\begin{equation} 
\vert \langle 0 \vert M_3 \vert 0\rangle \vert > 2. \label{M3viol}
\end{equation} 
As one can determine, the main difficulty which arises when dealing with the Mermin's inequalities is the presence of an increasing number of field operators $A,B,C, ...$, thus reflecting the multipartite nature of these inequalities. Unlike the Bell-CHSH inequality case, facing this point requires the introduction of several spectral subspaces of the modular operator $\delta$. More precisely, we shall make use of a sequence of orthonormal vectors $\phi_\nu$ 
\begin{align}
    \langle \phi_\mu \vert \phi_\nu \rangle = \delta_{\mu \nu}, \label{orth}
\end{align}
belonging to the spectral subspaces defined by
\begin{align}
I_\nu = \left[ \lambda_\nu^2 - \varepsilon, \lambda_\nu^2 +  \varepsilon  \right] \subset (0,1)  
\end{align}
where we have $\nu = 1, 2 , ...$ and $\lambda_\nu^2 \neq \lambda_\mu^2$ for $\nu \neq \mu$.

Concerning the test functions, we shall proceed by introducing the expressions 
\begin{align}
    f_\nu = \eta_\nu (1+s) \phi_\nu \;, \qquad  f'_\nu = \eta'_\nu (1+s) i\phi_\nu \;,\label{ffnu}
\end{align}
with $\phi_\nu$ as described above and $(\eta_\nu,\eta'_\nu)$  arbitrary real constants.  Since $s^2=1$, it turns out that we have $s f_\nu = f_\nu$ and $s f'_\nu = f'_\nu$. Let us consider now the action of the modular conjugation $j$. We have 
\begin{align}
    \delta (j \phi_\nu) = j (\delta^{-1} \phi_\nu) \;, \label{jmu}
\end{align}
showing that $j$ exchanges the subspace $I_\nu$ into 
\begin{align}
\tilde{I}_\nu = \left[1/ \lambda_\nu^2 - \varepsilon, 1/\lambda_\nu^2 +  \varepsilon  \right]. 
\end{align}
As a consequence,
\begin{align}
    \langle \phi_\mu \vert j \phi_\nu \rangle = 0,
\end{align}
since $\phi_\mu$ and $j\phi_\nu$  belong to different spectral subspaces.
Using the property $s f_\nu = f_\nu$, it follows that  $s^\dagger (j f_\nu) = j f_\nu$. In fact,
\begin{align}
    s^\dagger (j f_\nu) = j \delta^{-1/2} j  f_\nu = \delta^{1/2} f_\nu = j s f_\nu = j f_\nu.
\end{align}
We see, therefore, that  $f_\nu \in M$, while  $j f_\nu \in M'$. For $\mu \neq \nu$, we also have $\langle f_\mu \vert f_\nu \rangle = 0$, as well as $\langle j f_\mu \vert j f_\nu \rangle = \langle f_\mu \vert f_\nu \rangle = 0$. Similar properties hold for the primed versions $f'_\nu$. 

\begin{figure}[t!]
	\begin{minipage}[b]{1.0\linewidth}
		\includegraphics[width=\textwidth]{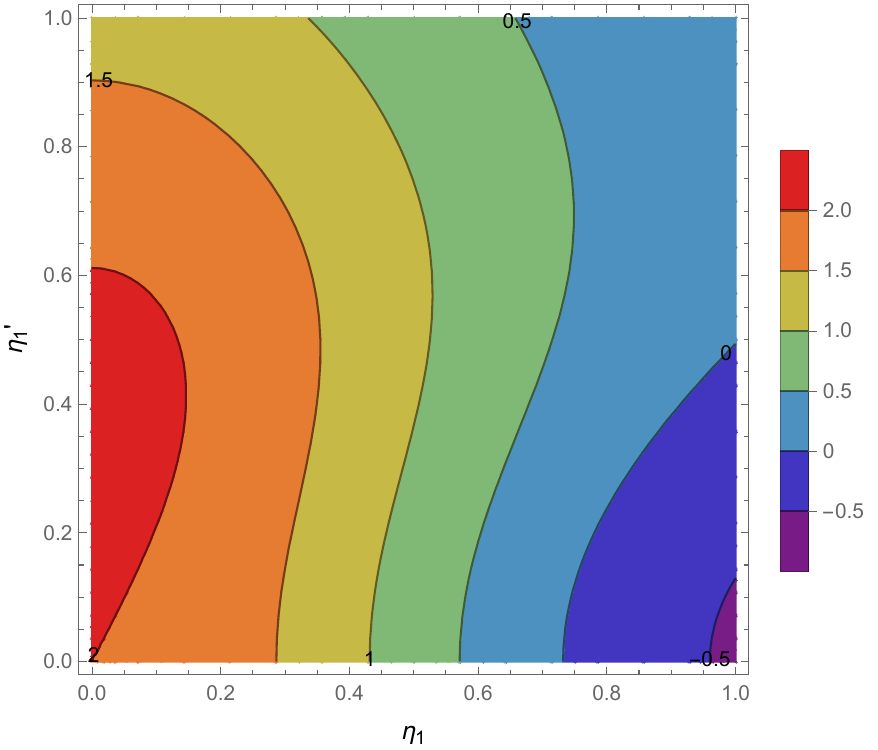}
	\end{minipage} \hfill
\caption{Contour plot exhibiting $\left \langle M_3 \right \rangle$ as a function of $\eta_1$ and $\eta_1'$ considering $\eta_1 = \eta_2$, $\eta_1' = \eta_2'$, $\lambda_1 = 0.9$, and $\lambda_2 = 0.6$. There is violation of Mermin's inequalities whenever  we have $\vert \left \langle M_3 \right \rangle \vert > 2$, that is, in the red region.}
	\label{lambda}
	\end{figure}

Employing the Weyl operators,  for $M_3$ we get 
\begin{align}
M_3 = W_{f'+g+h} + W_{f+g'+h} + W_{f+g+h'} - W_{f'+g'+h'}
\end{align}
where $(f,g,h)$ are generic test functions whose supports are spacelike separated from each other. Recalling that 
\begin{align}
\langle 0 \vert W_{f+g+h}  \vert 0 \rangle =  e^{-\frac{1}{2} \vert\vert f+g+h \vert\vert^2}, \label{nm3}
\end{align}
one gets 
\begin{align}
\langle  {M}_3 \rangle &=  e^{-\frac{1}{2} \vert\vert f'+g+h \vert\vert^2}   +  e^{-\frac{1}{2} \vert\vert f+g'+h \vert\vert^2} \nonumber \\
&+  e^{-\frac{1}{2} \vert\vert f+g+h' \vert\vert^2}   -  e^{-\frac{1}{2} \vert\vert f'+g'+h' \vert\vert^2}. \label{mm}
\end{align}
We still need to specify the test functions $(f,f')$, $(g,g')$, and $(h,h')$. To that end we shall make use of the expressions~\eqref{ffnu} and proceed in an alternate way, according to the following setup: 
\begin{itemize} 
\item The first pair of test functions, $(f,f')$ will be accommodated in the first spectral subspace $I_1$, namely, 
\begin{equation} 
f = f_1 \;, \qquad f'= f'_1 \;. \label{ftf}
\end{equation} 
\item The second pair $(g,g')$ will be obtained from the first pair, $(f_1, f'_1)$, by means of the modular conjugation $j$, 
\begin{equation} 
g = j f_1 \;, \qquad g'= j f'_1 \;. \label{stf}
\end{equation} 
\item The third pair $(h,h')$ will be accommodated in the second spectral subspace $I_2$, 
\begin{equation} 
h = f_2 \;, \qquad h'= f'_2 \;. \label{ttf}
\end{equation} 
\item The fourth pair will be obtained from the third one by acting upon with the modular operator $j$. 
\item The fifth pair will belong to $I_3$, while the sixth pair will be obtained from the fifth one through the operator $j$, and so on. 
\end{itemize} 
As one can easily verify, this procedure ensure that all pairs of test functions are spacelike separated. Therefore, taking into account that for $\langle M_3 \rangle$ we have only three pairs, it follows that 
\begin{align}
\langle f \vert h \rangle = \langle f \vert h' \rangle = \langle f' \vert h \rangle = \langle f' \vert h' \rangle &= 0. \label{lot1}
\end{align}
We can also find that
\begin{align}
    \vert\vert f \vert\vert^2 = \vert\vert jf \vert\vert^2 = {\eta_1}^2 (1+\lambda_1^2), \nonumber \\
    \vert\vert f' \vert\vert^2 = \vert\vert jf' \vert\vert^2 = {\eta'_1}^2 (1+\lambda_1^2),
\end{align}
and
\begin{align}
\vert\vert h \vert\vert^2 &= \eta_2^2 (1+ \lambda_2^2) \nonumber \\
\vert\vert h' \vert\vert^2 &= {\eta'_2}^2 (1+ \lambda_2^2).
\end{align}
Furthermore, it turns out that
\begin{align}
    \langle f \vert jf \rangle &= 2 \eta_1^2 \lambda_1, \nonumber \\
    \langle f' \vert jf' \rangle &= 2 {\eta'_1}^2 \lambda_1,
\end{align}
with all the other inner products relevant for this computation vanishing. Therefore, for the norms appearing in Eq.~\eqref{mm} we obtain: 
\begin{align}
    \vert\vert f'+ jf + h \vert\vert^2 &= {\eta'_1}^2 (1+\lambda_1^2)  + \eta_1^2 (1+\lambda_1^2)  + \eta_2^2(1+\lambda_2^2), \nonumber \\
    \vert\vert f+ jf' + h \vert\vert^2 &= \eta_1^2 (1+\lambda_1^2)  + {\eta'_1}'^2 (1+\lambda_1^2)  + \eta_2^2(1+\lambda_2^2), \nonumber \\
     \vert\vert f+ jf + h' \vert\vert^2 &=  2 \eta_1^2(1+\lambda_1^2 + 2 \lambda_1) + {\eta'_2}^2(1+\lambda_2^2), \nonumber \\
      \vert\vert f'+ jf' + h' \vert\vert^2 &=  2 {\eta'_1}^2(1+\lambda_1^2+2\lambda_1 ) + {\eta'_2}^2(1+\lambda_2^2).
\end{align}
Finally, the full expression for $\langle M_3 \rangle$ turns out to be
\begin{align}
\langle  {M}_3 \rangle &= 2 e^{-\frac{1}{2} \left( (\eta_1^2 + {\eta'_1}^2) (1+\lambda_1^2) + \eta_2^2(1+\lambda_2^2) \right)}  \nonumber \\
&+ e^{-\frac{1}{2} \left( 2\eta_1^2(1+\lambda_1)^2 + {\eta'_1}^2(1+\lambda_2^2) \right)}  \nonumber \\
&- e^{-\frac{1}{2} \left( 2{\eta'_1}^2(1+\lambda_1)^2 + {\eta'_2}^2(1+\lambda_2^2) \right) } \label{fff}
\end{align}
Using the above expression, one can search for Mermin inequality violations by scanning the available parameter space. It turns out that, exactly as in the Bell-CHSH inequality case~\cite{Weyl23}, Eq.~\eqref{fff} is able to capture the Mermin inequality violation at the quantum level, as one can see in Fig.~\ref{lambda}. 

\section{Discussing the use of normal operators}\label{N}

As already underlined, one interesting feature of the results presented in the previous sections is that the violation of both Bell-CHSH and Mermin inequalities has been obtained by making use of the Weyl operators, $ W_f = e^{i \phi(f) }$, which are not Hermitian operators. Rather, being unitary, they belong to a broader class of operators, namely, the so-called normal operators $\{ {\cal A} \}$,
\begin{equation} 
{\cal A} {\cal A}^{\dagger} = {\cal A}^{\dagger} {\cal A} \;. \label{normal}
\end{equation}
The results obtained thus far by making direct use of normal operators can be understood with the help of the following considerations: 
\begin{itemize} 
\item Even being unitary, the vacuum expectation value of the Weyl operators is real,  
\begin{equation} 
\langle W_f \rangle = e^{-\frac{1}{2} \vert \vert f \vert \vert^2}. \label{wreal}
\end{equation}
\item The standard classical argument based on the local realism generalizes to unitary complex numbers, {\it i.e.},
\begin{equation} 
\vert (z+z') w + (z-z') w'\vert \le 2  \;, \label{zzz}
\end{equation}
where $|z| = |z'|=|w| = |w'| =1$. As such, the Bell-CHSH and Mermin inequalities  have the meaning of  correlation functions 
built out from unitary operators which violate the classical bounds~\eqref{zzz}. 
\item Despite many efforts and many trials, we have been unable to find an explicit Hermitian combination of the Weyl operators leading to an explicit violation of Bell-CHSH and Mermin inequalities. To give an example of what is going on, the simplest Hermitian combination, $\frac{1}{2}(W_f + W_f^{\dagger})$ does not produce any violations. A very huge set of more sophisticated combinations has been examined without any success. The situation looks very similar to the current stage of the GHZ theorem~\cite{Erhard_2020}, for which no Hermitian operators have been found so far in order to generalize the theorem to the multidimensional case. 
\end{itemize} 

More generally, the potential use of normal operators in the study of the Bell-CHSH inequality has already been advocated in Refs.~\cite{Hu_2017,Erhard_2020}, where the authors outlined a series of suggestive and consistent features. 

To grasp the depth of this ongoing discussion, let us present some very simple reasoning, borrowed from~\cite{Hu_2017}. Write 
\begin{equation} 
{\cal A} = {\cal B} + i {\cal C}, \label{arg}
\end{equation}
with ${\cal B}$ and ${\cal C}$ self-adjoint. With $\cal A$ being a normal operator, it follows that 
\begin{equation} 
[{\cal A}, {\cal A}^\dagger]=0 \implies [{\cal B}, {\cal C}]=0. \label{argg}
\end{equation}
One sees thus that condition $[{\cal A}, {\cal A}^\dagger]=0$ implies that ${\cal A}$ is made up by two self-adjoint commuting operators. According to Quantum Mechanics, the operators $({\cal B}, {\cal C)}$ have a real spectrum and can be measured simultaneously. Therefore, a normal operator can be seen as arising from the simultaneous measurement of two commuting observables. Evidently, one has the freedom of collecting the two outcomes into a complex number. In particular, Weyl operators correspond to a special case of this general pattern. 

\vspace{-0.3cm}
\section{Conclusions}\label{C}
	
In this paper we investigated Mermin's inequalities within the framework of Quantum Field Theory, for the case of a free real scalar field. Using the powerful setup of Tomita-Takesaki modular theory and considering the von Neumann algebra built with Weyl operators, we were able to compute the Mermin correlation analytically and show that there is a region in the parameter space in which Mermin's inequalities are violated when probed in the vacuum state of a scalar field theory. We remark that the general construction presented here could be straightforwardly adapted to $n$-partite systems with $n>3$.  

It is worth underlining that the combination of the Algebraic Quantum Field Theory with the Tomita-Takesaki modular theory proved to be very adequate for a systematic study of both Bell-CHSH and Mermin inequalities. Moreover, the use of Weyl operators gave us the opportunity to address the very interesting ongoing discussion of the role that the class of normal operators might have at the quantum level.

Let us end by adding a few remarks on two challenging issues which we plan to address in the near feature. The first one is to face the Bell-CHSH and Mermin inequalities in the case of interacting Quantum Field Theories. The main goal here would be to evaluate the correlation function dependence on the coupling constant, analyzing whether the interaction increases or decreases the size of the Bell-CHSH and Mermin inequality violation in comparison with the free case. We plan to address this question first in the $1+1$ interacting Thirring model, which we already analyzed in the free case~\cite{Haar23}. As one can easily determine, the reason to study such a model relies on the fact that it has been solved exactly. As such, the K\"allén-Lehmann spectral density for the two-point correlation function of the fermion field is available in a closed form, enabling us to quantify the interaction effects. This is work in progress and we hope to report on this subject soon.
Though, it is worth mentioning that, by combining the Tomita-Takesaki modular theory with Unruh-DeWitt detectors, we were able to study the interaction between a scalar field and a pair of $q$-bits. Although unlike the case of a self-interacting Quantum Field Theory, the Bell-CHSH inequality has been evaluated in exact form. As reported very recently in~\cite{UDW_01_2024}, the presence of a scalar field induces a damping factor, thereby decreasing the size of the Bell-CHSH inequality violation in comparison with the case without it.

Another topic for future work is to consider the finite-temperature case and to inquire about the main differences in our results in this more realistic framework, allowing for a wider range of applications for our present results. Even restricting ourselves to the simpler case of free theories, the finite-temperature setting looks very attractive. As a very preliminary and intuitive guess, we can say that the existence of contributions in the dimensionless parameters $\left( LT,\; \frac{m}{T}\right)$ cannot be excluded, where $L$ is the separation distance between Alice's and Bob's regions, $T$ is the temperature, and $m$ is the quantum field mass. That would imply a more in-depth study of the correlation function's behavior in this finite-temperature scenario and will be reported  in future work.
	

\vspace{-0.3cm}
\section*{Acknowledgments}
\vspace{-0.2cm}
	The authors would like to thank the Brazilian agencies CNPq and CAPES, for financial support.  S.P.~Sorella, I.~Roditi, and M.S.~Guimaraes are CNPq researchers under contracts 301030/2019-7, 311876/2021-8, and 310049/2020-2, respectively.

	
\end{document}